\documentclass{article}

\usepackage{amsmath}
\usepackage{amssymb}
\usepackage{amsthm}
\usepackage{amsfonts}
\usepackage{array}
\usepackage{graphicx}
\usepackage{mathrsfs}
\usepackage{listings}
\usepackage{times}
\usepackage{bm}
\usepackage{natbib}
\usepackage{multirow}
\usepackage[plain,noend]{algorithm2e}
\usepackage[margin=0.7in]{geometry}

\usepackage{setspace}

\newtheorem{theorem}{Theorem}[section]

\newcommand{\ba}{\begin{array}}
\newcommand{\ea}{\end{array}}

\begin{document}

\markboth{Rui Zhang, Kwun Chuen Gary Chan}{Marginalizable mixed effects model for clustered binary data}

\title{Marginalizable conditional model for clustered ordinal data}

\author{Rui Zhang, Kwun Chuen Gary Chan\\
Department of Biostatistics, University of Washington, Seattle, WA, USA \\
zhangrui@uw.edu,kcgchan@uw.edu}

\maketitle

\begin{center}
ABSTRACT
\end{center}
We introduce a flexible parametric mixed effects model for correlated binary data, with parameters that can be directly interpreted as marginal odds ratios. This leads to a robust estimation equation with an optimal weighting matrix being the inverse of a genuine model-based covariance matrix. Flexible correlation structures can be imposed by correlated random effects, and correlation parameters can be estimated by solving a composite likelihood score function. Marginal parameters are consistently estimated even when the conditional parametric model is misspecified, and the robust estimation procedure has low estimation efficiency loss compared to the maximum likelihood estimation under a correct model specification. Simulations, analyses of the Madras longitudinal schizophrenia study and British social attributes panel survey were carried out to demonstrate our method.


\section{Introduction}

Correlation exists naturally when observations are grouped into clusters. For instance, observations are collected from the same subjects at different time points in longitudinal studies. For observations within a cluster, data are typically correlated even after adjusting for observed covariates. We need to address such correlations in a valid statistical analysis. One can often evaluate two distinct covariate effects from clustered data: the \textit{marginal} covariate effect as a population-averaged effect from the study population and the \textit{conditional} covariate effect that quantifies the effect conditional on some unobservable random effects, e.g. cluster-specific effects. Distinct models and methods have been proposed for estimating the marginal and conditional covariate effects. However, marginal and conditional models are typically incompatible for non-linear models such as a logistic regression model. In this paper we consider an unified marginal and conditional model for correlated binary data $Y$ and a vector of covariates $X$, with $(X_i,Y_i)$ being a vector of observations from the $i^{th}$ cluster and $(X_{ij}$, $Y_{ij})$ denoting the $j^{th}$ component/observation, $j=1,\ldots, n_i$. Generalization to three-level clustered data will be discussed in Section 4.

Marginal models were introduced to estimate the marginal covariate effects, which are directly interpretable and are preferred to answer public health questions, according to \cite{ref28} and \cite{ref7}. These models are often semi-parametric, which only assume the first and perhaps the second moments of outcomes conditioning on covariates. Under a marginal mean model $E(Y_{ij}\mid X_{ij})=g(X_{ij}^T\beta)$ where $g$ is a known inverse link function, the parameter $\beta$ represents a transformation of the population-average change in expected response per unit change in a given predictor, controlling for the other covariates. For $\beta$ inference, \cite{ref106} proposed the Generalized Estimating Equation (GEE). For a dataset containing $m$ independent clusters, the estimate is obtained by solving
\[
\underset{i=1}{\overset{m}{\sum}}D_i^TV_i^{-1}S_i=0\ ,
\]
where $S_i=Y_i-g(X_i^T\beta)$, $D_i=\partial g(X_i^T\beta)/\partial \beta$ and $V_i$ is a "working" covariance matrix given by $V_i=A_i^{1/2}R(\alpha)A_i^{1/2}/\phi$, $A_i$ is a diagonal matrix with elements proportional to $\mathrm{var}(Y_i)=h(X_i^T\beta)/\phi$ and $R(\alpha)$ is a cluster-common working correlation matrix parametrized by $\alpha$. Nuisance parameters $\alpha$ and $\phi$ are typically estimated by the method of moments.
\cite{book7} pointed out that the optimal estimation efficiency will be achieved when $V_i$ is the true covariance matrix of $Y_{i}$. Note that the working correlation matrix $R(\alpha)$ may not correspond to a genuine correlation matrix from any plausible joint distribution of binary outcomes, as discussed by \cite{ref101}, who argued that $R(\alpha)$ should be viewed as a weighting matrix, and $\alpha$ should be fixed instead of being estimated. 

On the other hand, mixed effect models are commonly used for modeling conditional covariate effects. In general, some unobservable random effects are introduced to model latent cluster effects that cannot be explained by observed covariates and thus together with covariates, they fully characterize correlations between observations; i.e., conditioning on random effects and covariates, observations are assumed to be independent. These models gained popularity because complex correlation structures can be modeled naturally by Gaussian random effects, cluster-specific predictions can be made and likelihood inference is directly applicable. Let $b_{ij}$ denote an unobserved random effect with a conditional density $f(b_{ij}\mid X_{ij})$. A conditional parametric model specifying the distribution of outcome given observed covariates $X_{ij}$ and random effects $b_{ij}$ is typically assumed. The observed likelihood can be constructed as a marginal density by integrating this conditional outcome density over the random effect distribution:
\begin{equation}\label{MixModel1}
\mathrm{pr} (Y_{ij}\mid X_{ij})=\int \mathrm{pr} (Y_{ij}\mid X_{ij},b_{ij})f (b_{ij}\mid X_{ij})db_{ij} \ .
\end{equation}
However, in general there does not exist a closed-form expression for \eqref{MixModel1} except for the Gaussian linear mixed model and a few other special cases, causing two problems: 1) observed data likelihood inference requires heavy computation; 2) we cannot directly estimate marginal covariate effects because the lack of a closed-form expression.

Regarding the first problem, numerical integration/approximation techniques have been developed to maximize the observed data likelihood or its approximations, such as the penalized quasi-likelihood inference by \cite{ref56}, Laplace approximations by \cite{ref19}, Gauss-Hermite quadrature and Monte Carlo importance sampling algorithms by \cite{ref24}. Related methods have been described in details and compared by \cite{ref17}. 

Several authors have offered solutions to the second problem from different perspectives. From \eqref{MixModel1}, we know that a marginal model and a random effect distribution will jointly determine a conditional model. Likewise, marginal and conditional models will jointly determine the random effect distribution. \cite{ref7} and \cite{ref12} first jointly modeled the marginal mean model and the random effect distribution and then solved for the conditional mean model, giving the marginalized multilevel models. While this method is conceptually appealing, the implementation is not straightforward since a deconvolution problem is involved, leading to a certain difficulty to the model formulation and interpretation. The bridge distribution proposed by \cite{ref15} started from a fixed pair of marginal and conditional mean models; the authors solved for the random effect distribution and named it the bridge distribution. However, the bridge distribution  may not correspond to any known parametric distribution and a lack physical interpretation is also a concern. One may model the joint distribution of a random vector from marginal distributions using a copula, and \cite{ref113} applied this approach for marginal reference.


Our model formulation for binary data starts from a different perspective. A conditional mean model and a family of correlated random effects are specified to complete the parametric specification of the joint distribution, while directly leading to a marginal logistic regression model. Our formulation is partly motivated from frailty models in survival analysis. The model formulation will be discussed in Section 2. Robust inference is developed in Section 3, extending the generalized estimating equation in \cite{ref106} and the alternating logistic regression proposed by \cite{ref6}. The marginal odds ratio parameters can be consistently estimated even when the working conditional mean model or the random effect distribution is misspecified. Asymptotic properties of the estimators are presented in Subsection 3$\cdot$4. In Section 4, we discuss extensions to three-level clustered data. We show the three-level correlation structure can be naturally incorporated into our model and thus marginal inference can be easily extended into this case. Numerical simulations will be presented in Subsection 5$\cdot$1, which demonstrate the proposed estimator has a small bias, is robust against model mis-specification and has a negligible efficiency loss compared to maximum likelihood inference. Analyses of the Madras longitudinal schizophrenia study and the British social attributes panel survey will be presented in Subsections 5$\cdot$2 and 5$\cdot$3. Concluding remarks and discussions will be given in Section 6. Technical conditions and a proof of the main theorem will be provided in the appendix.  

\section{A marginalizable conditional model for correlated binary data}
\subsection{A motivation from frailty models}
Our marginalizable mixed effect model is motivated from a close examination of Cox-type frailty models from the survival analysis literature. In these models, given values of frailty $a_{ij}$ and covariate $x_{ij}$, the conditional hazard rate at time $t$ of the $j^{th}$ observation from the $i^{th}$ cluster is formulated by $a_{ij}\lambda_0(t)exp(x_{ij}^T\beta)$, where $\lambda_0(\cdot)$ is an unspecified baseline hazard rate function. Its conditional survival probability is
\begin{equation}\label{frailty.conditional.model}
S(t\mid X_{ij},a_{ij})=\mathrm{exp}\left(-a_{ij}\Lambda_0(t)e^{X_{ij}^T\beta}\right),\quad\text{where } \Lambda_0(t):=\int^t_0\lambda_0(s)ds.
\end{equation}
The frailties $a_{ij}$ are equivalent to exponentiated random intercepts. For model identifiability in the presence of an unknown baseline hazard rate $\lambda_{0}(\cdot)$, no intercept term is included into the frailty models and one assumes $\mathrm{E}(a_{ij})=1$.

It is common to assume the frailty follows a Gamma distribution with mean one and unknown variance $1/\gamma$ to be estimated, with the density
\[
f_\gamma(a)=\frac{\gamma^\gamma}{\Gamma(\gamma)}a^{\gamma -1}e^{-\gamma a} \ ,
\]
see \cite{ref75}, \cite{ref93}, \cite{ref34}, \cite{ref37} and \cite{ref38} for relevant discussions.Integrating over $a_{ij}$ gives the marginal survival probability
\[
S(t\mid X_{ij})=\int^\infty_0 \mathrm{exp}\left(-a_{ij}\Lambda_0(t)e^{X_{ij}^T\beta}\right)\frac{\gamma^\gamma}{\Gamma(\gamma)}a_{ij}^{\gamma-1}e^{-\gamma a_{ij}}da_{ij}=\left(\frac{1}{1+\Lambda_0(t)e^{X_{ij}^T\beta-log \gamma}}\right)^\gamma \ .
\]
Setting $\gamma=1$, frailties become marginally exponential distributed and the above marginal survival probability simplifies into
\[
S(t\mid X_{ij})=\frac{1}{1+\Lambda_0(t)e^{X_{ij}^T\beta}}\ .
\]
Thus at $\gamma=1$, $\beta$ can be marginally interpreted as the log failure odds ratio.

\subsection{A model for correlated binary data}
In the absence of censoring and suppose we are interested in modeling the survival probability at a certain time point $t^*$, correlated survival outcomes are equivalent to correlated binary outcomes where the binary outcome is $Y_{ij}=I(T_{ij} > t^*)$, where $T_{ij}$ a survival outcome. We assume the conditional probability of the binary outcome follows
\begin{equation}\label{OurModel1}
\mathrm{pr}(Y_{ij}=1\mid X_{ij},a_{ij})=\exp\left(-a_{ij}e^{-X_{ij}^T\beta }\right)\ ,
\end{equation}
where $a_{ij}$'s are marginally standard exponential distributed. In this formulation an intercept is included into the linear predictor, corresponding to $\mathrm{log}\left\{\Lambda_0(t^*)\right\}$ from \eqref{frailty.conditional.model}.

Similarly to the survival model, the marginal survival probability becomes
\begin{equation}\label{marginal}
\mathrm{pr}(Y_{ij}=1\mid X_{ij})=\frac{1}{1+e^{-X_{ij}^T\beta}}=\frac{e^{X_{ij}^T\beta}}{1+e^{X_{ij}^T\beta}} \ .
\end{equation}
Therefore marginally, outcomes follow a logistic regression model with the same $\beta$ coefficients as in the working conditional model \eqref{OurModel1}. We describe the conditional model as a working model, because in Section 3 we will propose a robust estimator for $\beta$ under the marginal model \eqref{marginal}, which is consistent even when the working conditional model \eqref{OurModel1} is misspecified.

\subsection{Random effect variance}

Now suppose frailties are exponentially distributed with a variance $\gamma^{-2}$, a similar marginalization as in Subsection 2$\cdot$2 can be obtained, where
\[
\mathrm{pr}(Y_{ij}=1\mid X_{ij})=\frac{1}{\gamma}\int^\infty_0 \mathrm{exp}\left(-a_{ij}e^{-X_{ij}^T\beta} -\frac{a_{ij}}{\gamma} \right)da_{ij}=\frac{1}{\gamma}\frac{1}{e^{-X_{ij}^T\beta}+\frac{1}{\gamma}}=
\frac{1}{e^{-X_{ij}^T\beta+log\gamma}+1} \ .
\]
We can see $\mathrm{log}(\gamma)$ merges with the intercept in the marginal probability, implying $\gamma$ is not identifiable marginally. Moreover, $\gamma$ is not identifiable in the joint likelihood. For example,
\begin{eqnarray*}
&&\mathrm{pr}(Y_{1}=0,Y_{2}=1,\ldots,Y_{n}=1)
\\=&&\mathrm{pr}(Y_{2}=1,\ldots,Y_{n}=1)-\mathrm{pr}(Y_{1}=1,\ldots,Y_{n}=1)\\=&&|I+C_{-1}\text{diag}(
\gamma e^{-X_2^T\beta},\ldots,\gamma e^{-X_n^T\beta}
)|^{-1} -|I+C\text{diag}( \gamma e^{-X_1^T\beta},\ldots,\gamma e^{-X_n^T\beta})|^{-1}\ ,
\end{eqnarray*}
where $C_{-1}$ is the element-wise square root of the correlation matrix between $(a_2,\ldots,a_n)$ and $C$ is the element-wise square root of the correlation matrix between $(a_1,\ldots,a_n)$. Therefore, $\log(\gamma)$ merges with the intercept in joint probabilities as well, and the variance of the random effect cannot be separately estimated from the intercept.

In view of this identifiability problem, we will standardize the random effect distribution having a unit variance.

We note that in conventional linear and logistic mixed models, the within-cluster correlation is controlled by the variance of some shared random effects. 
While the variance of the random components is standardized in our model, flexible within-cluster correlation can still be modeled by correlated random effects as discussed below, as opposed to using a shared random effect that is often assumed in conventional models.

\subsection{Random effect correlation}
We allow the frailties to be correlated within clusters and follow a multivariate exponential distribution, instead of assuming frailties are identical within each cluster. To facilitate the modeling of correlations, a class of multivariate exponential distributions can be constructed from multivariate normal distributions as shown in \cite{ref29} and \cite{ref18}. Set $W_1$ and $W_2$ to be two independent $p$-variate, zero-mean and unit-variance Gaussian distributed random vectors; i.e. $W_j=(W_{j1},\ldots,W_{jp}), j=1,2$. Denote their $p\times p$ correlation matrix by $C$. Let $Z_k=(W_{1k}^2+W_{2k}^2)/2, k=1,\ldots,p$.  For each $k$, $2Z_k$ is marginally $\chi^2(2)$ distributed; therefore $Z_k$ follows a standard exponential distribution. Moreover, the correlation matrix $R$ of the random vector $(Z_1,\ldots,Z_p)$ is an element-wise square of $C$, see \cite{ref18} for related discussions.

The above connection between multivariate exponential and Gaussian distributions allows one to model flexible correlation patterns similar to the Gaussian mixed effect models. In the following, we will parametrize the correlation matrix for a multivariate exponential random vector by a possibly vector-valued parameter $\rho$ and we will discuss models for three-level clustered data in Section 4.

\subsection{Generalization to covariate-dependent distributed frailties}
Although we originally considered the frailty distribution to be independent of covariates, the proposed method for marginal inference is unaffected when covariates are covariate-dependent.  Suppose given a covariate $X_{ij}$, $a_{ij}$ is exponential distributed with mean $e^{X_{ij}^T\gamma}$, and a frailty vector $a_i$ given $X_i$ has a correlation matrix $R$. Consider the rescaled frailty vector $\tilde{a}_{i}:=(e^{-X_{i1}^T\gamma}a_{i1},\ldots,e^{-X_{in_i}^T\gamma}a_{in_i})$, which is multivariate exponential with mean one and has the same correlation matrix $R$, the conditional probability of the binary outcome follows from \eqref{OurModel1}, i.e.
\[
\mathrm{pr}(Y_{ij}=1\mid X_{ij},a_{ij})=\mathrm{exp}\left(-a_{ij}e^{-X_{ij}^T\beta}\right)
=\mathrm{exp}\left(-\tilde{a}_{ij}e^{-X_{ij}^T\tilde{\beta}}\right), \quad\text{where }\tilde{\beta}=\beta-\gamma\ .
\]
And the marginal probability \eqref{marginal} becomes
\[
\mathrm{pr}(Y_{ij}=1\mid X_{ij})=\frac{e^{X_{ij}^T\tilde{\beta}}}{1+e^{X_{ij}^T\tilde{\beta}}}\ .
\]
For marginal inference, the parameter of interest is $\tilde{\beta}$ and can be estimated as if the frailties were covariate independent.
\section{Estimation}
\subsection{Estimating equation for $\beta$ with an optimal weighting matrix.}

Since the proposed model is parametric, it is natural to consider maximum likelihood estimation (MLE) for model inference, as discussed by \cite{ref30} and \cite{ref22}. However, MLE has two major drawbacks. First, obtaining consistent MLE requires a correct specification of the conditional model and the random effect distribution, even when the marginal parameters are of main interest. Besides, the likelihood function involves up to $2^{n}-1$ terms for each cluster, where $n$ is the cluster size. It may be practically infeasible to compute MLE even for a moderate cluster size, since the computation burden grows exponentially with cluster size.

We propose a robust estimation procedure for the marginal covariate effects $\beta$, by replacing the working correlation matrix $R(\alpha)$ in Liang and Zeger's GEE with a real correlation matrix, derived from the conditional model in \eqref{OurModel1}.

Denote the whole set of parameters by $\theta:=(\beta,\rho)$. Let $g$ be the inverse of the logit link function:
\[
g(x_{ij}^T\beta):=\mathrm{pr}(Y_{ij}=1\mid  X_{ij}=x_{ij})=\exp(x_{ij}^T\beta)/(1+ \exp(x_{ij}^T\beta)) \ .
\]
For $\beta$ inference, we solve for
\begin{equation}\label{EE1}
\frac{1}{m}\underset{i=1}{\overset{m}{\sum}}D(X_i;\beta)^T V^{-1}(X_i;\theta)S(X_i,Y_i;\beta)=0\ ,
\end{equation}
where $D(X_i;\beta)=\partial g(X_{i}^T\beta)/\partial \beta$, $S(X_i,Y_i;\beta)=Y_i-g(X_{i}^T\beta)$ and $V(X_i;\theta)$ is the $n_i\times n_i$ covariance matrix of the outcome $Y_i$. To be more specific,
the $j^{th}$ diagonal entry of $V(X_i;\theta)$ is given by
\[
V_{jj}(X_i;\theta)=\frac{e^{x_{ij}^T \beta}}{\left(1+e^{x_{ij}^T \beta}\right)^2}\ .
\]
Its $j^{th}$ row and $k^{th}$ column entry is 
\[
V_{jk}(X_i;\theta)=\left[\frac{1}{(1-\rho_{jk})e^{-(x_{ij} +x_{ik})^T\beta}+e^{-x_{ij}^T\beta} +e^{-x_{ik}^T\beta}+1}-\frac{1}{1+e^{-x_{ij}^T\beta}}\frac{1}{1+e^{-x_{ik}^T\beta}}\right]\ ,\quad j\neq k \ ,
\]
where $\rho_{jk}$ is the correlation between $a_{ij}$ and $a_{ik}$. In the case of an exchangeable correlation structure, $\rho_{ij}$ are identically equal to a scalar $\rho$. In the case of an auto-regressive with degree one correlation structure, $\rho_{jk}$ are functions of a scalar parameter $\rho$. In more general correlation structures, such as the un-structured correlation structure, $\rho_{ij}$ are functions of some vector-valued parameter $\rho$.

For the estimating equation in \eqref{EE1}, any plug-in value of $\rho$ between 0 and 1 will give a consistent estimate of $\beta$. When the true value $\rho_0$ or a consistent estimate of $\rho_0$ is plugged into \eqref{EE1}, the estimate of $\beta$ is consistent and efficient within a class of linear estimating equations, as long as the marginal model is correct, according to \cite{book4}. An estimator of $\rho$ is given in the next subsection, and theorems justifying the above remarks will be given in Subsection 3$\cdot$4.

\subsection{Estimating $\rho$ via composite likelihood.}
Concerned with the computation burden discussed before, we choose to maximize a composite likelihood function over $\rho$ with a fixed $\beta$. The composite likelihood for a single cluster is just the summation of all pairwise likelihoods. Denote
\begin{eqnarray*}
p_{ij}&=&\mathrm{pr}(Y_{ij}=1\mid X_{ij}=x_{ij})=g(x_{ij}^T\beta)\ ,\\ 
p_{ijk}&=&\mathrm{pr}(Y_{ij}=Y_{ik}=1\mid X_{ij}=x_{ij},X_{ik}=x_{ik})=\left[(1-\rho_{jk})e^{-( x_{ij}+x_{ik})^T\beta}+e^{- x_{ij}^T\beta}+e^{- x^T_{ik}\beta}+1\right]^{-1} \ .
\end{eqnarray*}
For a dataset containing $m$ independent clusters, the composite log-likelihood is defined as
\begin{eqnarray*}
&&\frac{1}{m}\sum_{i=1}^m\sum_{j<k}l_{jk}(X_i,Y_i;\theta)\\
&=&\frac{1}{m}\sum_{i=1}^m\sum_{j<k}\Big( y_{ij}y_{ik}\mathrm{log} p_{ijk} +(1-y_{ij})y_{ik}\mathrm{log} (p_{ik}-p_{ijk})\\&&\quad\quad\quad+ y_{ij}(1-y_{ik})\mathrm{log} (p_{ij}-p_{ijk})
+ (1-y_{ij})(1-y_{ik})\mathrm{log} (1-p_{ij}-p_{ik}+p_{ijk})
\Big)\ .
\end{eqnarray*}
To estimate $\rho$, we solve for the equation
\begin{equation}\label{EE2}
\frac{1}{m}\sum_{i=1}^m\sum_{j<k}\frac{\partial l_{jk}}{\partial \rho}(X_i,Y_i;\theta)=0 \ .
\end{equation}
To estimate $\beta$ and $\rho$ jointly, \cite{ref14} suggested alternating between solving \eqref{EE1} with a fixed plug-in $\rho$ from \eqref{EE2}, and solving \eqref{EE2} with a fixed plug-in $\beta$ from \eqref{EE1}, until convergence, obtaining estimates ($\hat{\beta}_m,\hat{\rho}_m)$. We write $m$ to indicate an estimate based on a dataset containing $m$ independent clusters. This method can be viewed as a generalization of alternating logistic regression proposed by \cite{ref6}. 

\subsection{Simplification of estimation procedure}
We can reduce the above alternating estimation procedure of $(\beta,\rho)$ into four steps:

\noindent Step 1. Solving \eqref{EE1} with a fixed parameter $\rho_1$, obtaining $\hat{\beta}_{1m}$;

\noindent Step 2. Solving \eqref{EE2} with the plug-in $\hat{\beta}_{1m}$, obtaining $\hat{\rho}_{2m}$;

\noindent Step 3. Solving \eqref{EE1} with the plug-in $\hat{\rho}_{2m}$, obtaining $\hat{\beta}_{2m}$;

\noindent Step 4. Solving \eqref{EE2} with plug-in $\hat{\beta}_{2m}$, obtaining $\hat{\rho}_{3m}$.

The final estimate is $(\hat{\beta}_{2m},\hat{\rho}_{3m})$. This simplified procedure gives an asymptotically equivalent estimate of $\theta$ as the alternating solution of \eqref{EE1} and \eqref{EE2}, under a correct model specification. In the following we give a heuristic justification. A detailed proof is given in the first author's Ph.D dissertation (Zhang, 2014).

In Step 1, $\hat{\beta}_{1m}$ is a consistent estimator for $\beta_0$, due to the robustness of \eqref{EE1}; yet it is not efficient since $\rho_1$ is not necessarily the true value $\rho_0$, nor a consistent estimate of $\rho_0$. With a consistent estimator of $\beta$ plugged into \eqref{EE2}, $\hat{\rho}_{2m}$ is a consistent estimate of $\rho_0$ in Step 2. Then $\hat{\beta}_{2m}$ in Step 3 is a consistent and efficient estimate for $\beta_0$, and $\hat{\rho}_{3m}$ in Step 4 is a consistent estimate for $\rho_0$ and is asymptotically equivalent to the joint solution of \eqref{EE1} and \eqref{EE2}.

\subsection{Large sample properties}
In this section we provide several theories for the asymptotic behaviour of our estimator $(\hat{\beta}_{m},\hat{\rho}_{m})$. 
\begin{theorem}\label{th1}
Suppose conditions C1 $\sim$ C6 stated in the appendix are satisfied, then when $m\rightarrow\infty$,

\noindent (a) the solution $\hat{\theta}_m=(\hat{\beta}_m,\hat{\rho}_m)$ of equations in \eqref{EE1} and \eqref{EE2} is consistent for $\theta_0$;

\noindent (b) $\sqrt{m}\left\{(\hat{\beta}_m-\beta_0)^T,(\hat{\rho}_m-\rho_0)^T\right\}^T$ converges weakly to a normal distribution of mean zero and a covariance matrix $V$ given by
\[
V=\left\{\mathrm{E} (B)\right\}^{-1}\left\{\mathrm{E} (C)\right\}\left\{\mathrm{E} (B)^T\right\}^{-1}\ ,
\]
where
\begin{eqnarray*}
B&=&\left(\ba{cc}D(X;\beta_0)^TV^{-1}(X;\theta_0)D(X;\beta_0)&0\\ -{\underset{j<k}{\sum}}\frac{\partial^2 l_{jk}}{\partial \beta \partial\rho}(X,Y;\theta)\mid_{\theta_0}  & -{\underset{j<k}{\sum}}\frac{\partial^2 l_{jk}}{ \partial\rho^2}(X,Y;\theta)\mid_{\theta_0}\ea\right)\ ,\\
C&=&\left(\ba{c}D(X;\beta_0)^TV^{-1}(X;\theta_0)S(X,Y;\beta_0)\\ {\underset{j<k}{\sum}}\frac{\partial l_{jk}}{ \partial\rho}(X,Y;\theta)\mid_{\theta_0} \ea\right)^{\otimes 2}\ .
\end{eqnarray*}
\end{theorem}
Its proof can be found in the appendix.

The next theorem is for a misspecified conditional mean model or a misspecified random effect distribution but a correct marginal mean model.

\begin{theorem}\label{th2}
Suppose only the marginal mean model \eqref{marginal} is true, and all the other conditions in Theorem 1 are satisfied, then when $m\rightarrow\infty$,

\noindent (a) the solution $\hat{\theta}_m=(\hat{\beta}_m,\hat{\rho}_m)$ of equations \eqref{EE1} and \eqref{EE2} is consistent for $(\beta_0,\rho_1)$, where $\rho_1$ is the value that minimizing a Kullback-Leibler distance defined on composite likelihoods between the misspecified pairwise joint model and the true pairwise joint model: 
\[
KL_{\text{composite}}(L,L^*)=\mathrm{E}_0\left[ log \left\{\frac{{\underset{j<k}{\prod}}L(X_j,X_k,Y_j,Y_k;\beta_0,\eta)}{{\underset{j<k}{\prod}}L^*(X_j,X_k,Y_j,Y_k;\beta_0,\rho_1)}\right\}\right]\ ,
\]
where $L$ denotes the likelihood of the true pairwise joint model, $L^*$ for the mis-specified one, and $\eta$ is some other parameters under the true model.

\noindent (b) $\sqrt{m}\left\{(\hat{\beta}_m-\beta_0)^T,(\hat{\rho}_m-\rho_1)^T\right\}^T$
converges weakly to a normal distribution of mean zero and a covariance matrix $W$ given by
\[
W=\left\{\mathrm{E} (B_1)\right\}^{-1}\left\{\mathrm{E} (C_1)\right\}\left\{\mathrm{E} (B_1)^T\right\}^{-1}\ ,
\]
where
\[
B_1=\left(\ba{cc}D(X;\beta_0)^TV^{-1}(X;\beta_0,\rho_1)D(X;\beta_0)&0\\- {\underset{j<k}{\sum}}\frac{\partial^2 l^*_{jk}}{\partial \beta \partial\rho}(X,Y;\theta)\mid_{(\beta_0,\rho_1)}  & -{\underset{j<k}{\sum}}\frac{\partial^2 l^*_{jk}}{ \partial\rho^2}(X,Y;\theta)\mid_{(\beta_0,\rho_1)}\ea\right)\ ,
\]
\[
C_1=\left(\ba{c}D(X;\beta_0)^TV^{-1}(X;\beta_0,\rho_1)S(X,Y;\beta_0)\\ {\underset{j<k}{\sum}}\frac{\partial l^*_{jk}}{ \partial\rho}(X,Y;\theta)\mid_{(\beta_0,\rho_1)} \ea\right)^{\otimes 2}\ .
\]
\end{theorem}
As suggested in Theorem 1, when the pairwise conditional model is correct, the asymptotic covariance of $\sqrt{m}(\hat{\beta}_m-\beta_0)$ can be estimated by
\[
\hat{V}^\beta_{m}:=m\left(\underset{i=1}{\overset{m}{\sum}}
D(X_i;\hat{\beta}_m)^TV^{-1}(X_i;\hat{\theta}_m)D(X_i;\hat{\beta}_m)
\right)^{-1}\ .
\]
Allowing for a potentially mis-specified conditional model, a robust estimate of the asymptotic covariance of $\sqrt{m}(\hat{\beta}_m-\beta_0)$ is
\begin{eqnarray*}
\hat{V}^{\text{robust}}_{m}:=&&m\left(\underset{i=1}{\overset{m}{\sum}}
D(X_i;\hat{\beta}_m)^TV^{-1}(X_i;\hat{\theta}_m)D(X_i;\hat{\beta}_m)
\right)^{-1}
\left(\underset{i=1}{\overset{m}{\sum}}\left[ D(X_i;\hat{\beta}_m)^TV^{-1}(X_i;\hat{\theta}_m)S(X_i,Y_i;\hat{\beta}_m)\right]^{\otimes 2}\right)\\ && \cdot\quad
\left(\underset{i=1}{\overset{m}{\sum}}
D(X_i;\hat{\beta}_m)^TV^{-1}(X_i;\hat{\theta}_m)D(X_i;\hat{\beta}_m)
\right)^{-1}\ .
\end{eqnarray*}

\subsection{Discussion of inference methods and further remarks}
Inference by estimating equations \eqref{EE1} and \eqref{EE2} reduces the computation burden to $n_i^2$ for every cluster, compared to the order of $2^{n_i}$ in maximum likelihood inference. Alternative inference procedures may be adopted for the estimation of $\rho$; an example is the second-order GEE in \cite{ref21}. However, the computational burden of that method is in the order of $O(n_i^6)$, since it computes the inverse of a $n_i^2\times n_i^2$ matrix, which is the weighting matrix for pairwise outcome products. 

Under misspecification of the conditional distribution or the random effect distribution, the estimating equation \eqref{EE1} still guarantees consistency of the marginal parameter $\beta$, while the inverse weighting matrix $V$ is still a genuine covariance matrix, but corresponds to a misspecified model.

\section{Generalization to three-level clustered data}

For notational simplicity, our earlier discussions focused on two-level clustered data. Since our proposed model allows for flexible modeling of correlations between individual observations similar to Gaussian mixed effect models, it can be readily extended to datasets with a higher level of clustering. In this section, we consider a three-level clustered data where the first level consists of multiple independent clusters, inside each nested multiple individuals representing the second level, and multiple observations taken on every individual form the third level. Observations from different clusters are independent. Data from the $i^{th}$ cluster can be denoted by $(X_i,Y_i)=\mathrm{vec}(X_{ijk},Y_{ijk}):\, j=1,\ldots,n_i$ indexes individuals from the $i^{th}$ cluster and $k=1,\ldots,n_{ij}$ counts observations on the $j^{th}$ individual from the $i^{th}$ cluster.

We assume a similar working conditional model:
\[
\mathrm{pr}(Y_{ijk}=1\mid X_{ijk},a_{ijk})=exp\left(-a_{ijk}e^{-X_{ijk}^T\beta}\right),\quad a_{ijk}\sim \text{Exp}(1) \ .
\]
It is easy to show that the marginalization property of the working model still holds in the case of three-level clustering data:
\[
\mathrm{pr}(Y_{ijk}=1\mid X_{ijk}=x_{ijk})=\frac{e^{x_{ijk}^T\beta}}{1+ e^{x_{ijk}^T\beta}} \ .
\]
One way to model correlations among $a_{ijk}$'s is to assume that the level-two observations are exchangeable, and the level-three observations nested within level-two are also exchangeable. To be specific, we can model the correlations as follows:
\begin{eqnarray}
\mathrm{cor}(a_{ijk},a_{ij'k'})&=&\rho_{2}\ ,\quad j\neq j'\label{level3.1}\\
\mathrm{cor}(a_{ijk},a_{ij'k'})&=&\rho_{2}+ \rho_{3}\ ,\quad j=j'\ ,\quad k\neq k'\ .\label{level3.2}
\end{eqnarray}
Similar robust estimation methods based on \eqref{EE1} and \eqref{EE2} can still be used in this case. Denote $N_i=\sum_
{j=1}^{n_i}n_{ij}$ being the total number of observations from cluster $i$. For notational simplicity, we concatenate level-two observations in the cluster and denote $(X_i,Y_i)=\{\mathrm{vec}(X_{s},Y_{s}):s=1,\ldots,N_i\}$; i.e., we merge the double index $jk$ into a single index $s$. Suppose distinct observations $s_1,s_2$ are from individuals $j_1,j_2$ in the $i^{th}$ cluster respectively, then $\sum_{j=1}^{j_l-1}n_{ij}<s_l\leq \sum_{j=1}^{j_l}n_{ij}$, $l=1,2$.

Entries of the covariance matrix $V(X_i,\beta,\rho)$ are given by:
\[
V_{s_1s_1}(X_i;\beta,\rho)=\frac{e^{-X_{is_1}^T\beta}}{
\left( 1+e^{-X_{is_1}^T\beta}\right)^2}\ ,
\]
\begin{eqnarray*}
&&V_{s_1s_2}(X_i;\beta,\rho) \\=&&\frac{1}{\left\{1-\mathrm{cor}(a_{is_1},a_{is_2})\right\}e^{-(X_{is_1}+X_{is_2})^T\beta}+e^{-X_{is_1}^T\beta}+e^{-X_{is_2}^T\beta}+1}-\frac{1}{1+e^{-X_{is_1}^T\beta}}\frac{1}{1+e^{-X_{is_2}^T\beta}}\ .
\end{eqnarray*}
If we follow the exchangeable correlation formulation in \eqref{level3.1} and \eqref{level3.2},
\begin{small}
\begin{eqnarray*}
&&V_{s_1s_2}(X_i;\beta,\rho)\\=&&\left\{\ba{cc}\left\{(1-\rho_2-\rho_3)e^{-(X_{is_1}+X_{is_2})^T\beta}+e^{-X_{is_1}^T\beta}+e^{-X_{is_2}^T\beta}+1\right\}^{-1}-\left\{\left(e^{-X_{is_1}^T\beta}+1\right)\left(e^{-X_{is_2}^T\beta}+1\right)\right\}^{-1}\ ,&j_1=j_2\\
\left\{(1-\rho_2)e^{-(X_{is_1}+X_{is_2})^T\beta}+e^{-X_{is_1}^T\beta}+e^{-X_{is_2}^T\beta}+1\right\}^{-1}-\left\{\left(e^{-X_{is_1}^T\beta}+1\right)\left(e^{-X_{is_2}^T\beta}+1\right)\right\}^{-1}\ ,&j_1\neq j_2\ .
\ea\right.
\end{eqnarray*}
\end{small}
Similar to \eqref{EE2}, we write
\begin{eqnarray*}
&&\sum_{i=1}^m\sum_{s_1<s_2}l_{s_1s_2}(X_i,Y_i,\theta)\\
&=&\sum_{i=1}^m\sum_{s_1<s_2} \left\{y_{is_1}y_{is_2}\mathrm{log} p_{is_1s_2} +(1-y_{is_1})y_{is_2}\mathrm{log} (p_{is_2}-p_{is_1s_2})\right.\nonumber\\&&\left.\quad \quad \quad +y_{is_1}(1-y_{is_2})\mathrm{log} (p_{is_1}-p_{is_1s_2})+(1-y_{is_1})(1-y_{is_2})\mathrm{log} (1-p_{is_1}-p_{is_2}+p_{is_1s_2})\right\}\ ,
\end{eqnarray*}
where $p_{is_1}=\left(1+e^{-X_{is_1}^T\beta}\right)^{-1}$ and
\begin{small}
\[
p_{is_1s_2}=\left\{\ba{cc}\left\{(1-\rho_2-\rho_3)e^{-(X_{is_1}+X_{is_2})^T\beta}+e^{-X_{is_1}^T\beta}+e^{-X_{is_2}^T\beta}+1\right\}^{-1}\ ,&j_1=j_2\ ,\\
\left\{(1-\rho_2)e^{-(X_{is_1}+X_{is_2})^T\beta}+e^{-X_{is_1}^T\beta}+e^{-X_{is_2}^T\beta}+1\right\}^{-1}\ ,&j_1\neq j_2\ .
\ea\right.
\]
\end{small}
Similar to the case of two-level clustering, estimates are obtained by solving
\begin{eqnarray*}
\left\{\ba{c}\frac{1}{m}\underset{i=1}{\overset{m}{\sum}}D(X_i;\beta) V^{-1}(X_i;\beta,\rho)S(X_i,Y_i;\beta)=0\ ,\\
\frac{1}{m}\sum_{i=1}^m\sum_{s_1<s_2}\frac{\partial l_{s_1s_2}}{\partial \rho}(X_i,Y_i;\beta,\rho)=0 \ .\ea\right.
\end{eqnarray*}
Other correlation structures can also be used. For example, suppose the level-two observations are exchangeable units and the level-three observations are auto-regressive with order one, then we could model
\begin{eqnarray*}
\mathrm{cor}(a_{is_1},a_{is_2})&=&\rho_{2}\ ,\quad j_1\neq j_2\ ,\\
\mathrm{cor}(a_{is_1},a_{is_2})&=&\rho_{2}+ \rho_{3}^{|s_1-s_2|}\ ,\quad j_1=j_2\ .
\end{eqnarray*}
Entries in the inverse weighting matrix for estimating $\beta$ can be written as
\begin{eqnarray*}
&&V_{s_1s_2}(X_i;\beta,\rho) \\=&&\frac{1}{\left\{1-\mathrm{cor}(a_{is_1},a_{is_2})\right\}e^{-(X_{is_1}+X_{is_2})^T\beta}+e^{-X_{is_1}^T\beta}+e^{-X_{is_2}^T\beta}+1}-\frac{1}{1+e^{-X_{is_1}^T\beta}}\frac{1}{1+e^{-X_{is_2}^T\beta}}\ .
\end{eqnarray*}
We can write
\begin{small}
\begin{eqnarray*}
&&V_{s_1s_2}(X_i;\beta,\rho)\\ =&&\left\{\ba{cc}\left\{(1-\rho_2-\rho_3^{|s_1-s_2|})e^{-(X_{is_1}+X_{is_2})^T\beta}+e^{-X_{is_1}^T\beta}+e^{-X_{is_2}^T\beta}+1\right\}^{-1}-\left\{\left(e^{-X_{is_1}^T\beta}+1\right)\left(e^{-X_{is_2}^T\beta}+1\right)\right\}^{-1}\ ,&j_1=j_2\ ,\\
\left\{(1-\rho_2)e^{-(X_{is_1}+X_{is_2})^T\beta}+e^{-X_{is_1}^T\beta}+e^{-X_{is_2}^T\beta}+1\right\}^{-1}-\left\{\left(e^{-X_{is_1}^T\beta}+1\right)\left(e^{-X_{is_2}^T\beta}+1\right)\right\}^{-1}\ ,&j_1\neq j_2\ .
\ea\right.
\end{eqnarray*}
\end{small}
The four-step iterative estimation in Subsection 3$\cdot$3 still applies to this setting. 

\section{Numerical Studies}
\subsection{Simulation}
We conducted simulation studies to evaluate the finite sample performance of our proposed estimators. In each simulation scenario, $1000$ Monte Carlo datasets were generated. In each dataset, we generated 200 independent clusters. A covariate $X_1$ is included, which was a continuous normal random variable with mean zero and standard deviation $2$. 

Throughout this subsection, the marginal model was assumed to be
\begin{equation}\label{binary_simu}
\mathrm{pr}(Y_{ij}=1\mid X_{ij})=\frac{1}{1+exp(-\beta_0-\beta_1 X_{ij1})} \ ,
\end{equation}
where $\beta_0=1$ and $\beta_1=-1.2$. Under scenarios with joint distributions complying to our proposed models, we generated frailties from a multivariate standard exponential distribution with varying correlation structures, by the procedure discussed in Subsection 2$\cdot$4. To be specific, for the cases of two-level clustering, in which cluster sizes varied from $5$ to $7$ with equal probabilities, we imposed an exchangeable correlation structure and an auto-regressive of order one correlation structure. Exchangeable correlation structure is typically implemented to model correlations between individuals sampled from the same geographical region, hospital, etc; auto-regressive correlation usually models longitudinal observations over time. For the case of three-level clustering, we put $2$ or $3$ individuals into each cluster with probabilities $4/5$ and $1/5$ and generated $2$ or $3$ observations for each individual with probabilities $4/5$ and $1/5$. We imposed exchangeable correlation structures for both levels of clustering as discussed in Section 4. For model inference, we assumed the correct joint model and only model-based standard errors and 95\% confidence interval coverage rates were listed since their robust counterparts behaved quite similarly. 

A misspecified joint model was also considered, in which correlation was introduced via a latent variable model. For each cluster $i$, we generated an uniform variable $U_i$ and transformed it into a logistic distributed random variable $A_i=\mathrm{log} U_i -\mathrm{log}(1-U_i)$; at the end, we simulated $(Y_{i1},\ldots,Y_{in_i})$ by $Y_{ij}=I(X_{ij}^T\beta+A_i >0 )$, which satisfies the marginal model in \eqref{binary_simu}. For the proposed inference method, both the model-based and the robust standard errors and their respective 95\% confidence interval coverage rates were presented. 

Table 1 lists the simulation results in the case of two-level clustering under an exchangeable correlation structure and an auto-regressive of order one correlation structure, respectively in (a) and (b), under correctly specified joint models. The estimation efficiencies of our estimates, measured by mean squared error (MSE), are quite close to the MLE's, but the proposed method takes much less computing time than MLE. When $\rho=0.9$, $\beta$ estimate from MLE has a much larger bias compared to the proposed inference method.

Table 2 lists simulation results for three-level clustering. When the correlation is small, results from the two inference methods are pretty close. Otherwise, MLE estimates of $\beta$ are more biased. Besides, MLE behaves much worse than the proposed method in estimating the correlation parameters even when the correlation level is mild.

Table 3 lists simulation results for the mis-specified conditional model case. As expected, MLE of $\beta$ is biased while the proposed method gives consistent estimates of $\beta$, along with consistently estimated robust standard errors. 

\subsection{Madras longitudinal schizophrenia study}
We further demonstrate our proposed method using the Madras longitudinal schizophrenia study from \cite{ref26}, in which first-episode schizophrenics were followed for 10 years with the primary objective of characterizing the natural history of disease progression. The data contain several longitudinal binary outcome measurements indicating the presence of positive psychiatric symptoms over the time course: $t_{ij} = 0,\ldots , 11$ months during the first year following an initial hospitalization for $86$ schizophrenia patients. The binary outcome $Y_{ij}$ under interest is an indicator of whether or not a patient is observed to have thought disorders. Covariates include the time variable $t_{ij}$, a binary indicator $X_{ij2}$ of whether or not a patient is younger than 20 at disease onset and gender $X_{ij3}$: 0 for male and 1 for female. To assess the association between occurrence of thought disorders and the covariates, a marginal logistic regression model is constructed using a linear trend in time, with the time-independent binary covariates $X_{ij2}$ and $X_{ij3}$:
\[
logit \mathrm{ E}(Y_{ij}\mid t_{ij},X_{ij})=\beta_0 + \beta_1t_{ij}+ \beta_2X_{ij2}+ \beta_3X_{ij3} \ .
\]
Our regression model is almost identical to the model from \cite{ref7}, except that we did not center the time covariate.

Using the proposed method, we can answer whether the population-averaged probability of thought disorders differs across time, age-at-onset and gender subgroups. We used our proposed method and maximum likelihood to analyze this dataset, assuming observations from the same patients are exchangeable and auto-regressive with order one over time, i.e. AR(1). The results are reported in Table 4.
We can see the results from different inference methods are pretty similar, and the length of 95\% confidence intervals based on the proposed method is similar to those based on MLE. Since this is a longitudinal dataset, auto-regressive of order one (in time unit) correlation structure should be more close to the real situation, and in the following we report the results from our proposed inference method.

The estimated odds of thought disorder prevalence for a patient younger than 20 at the beginning of hospitalization is 47\% higher (95\% C.I.: 19\% lower to 166\% higher) than elder patient, controlling for gender and observation time. The estimated odds of thought disorder prevalence for a female patient is 46\% lower (95\% CI: 70\% lower to 4\% lower) than a male patient, controlling for age at onset and observation time. The estimated odds of thought disorder decreases by 29\% (95\% CI: 33\% to 24\%) in one month during hospitalization, controlling for age at onset and gender. There is evidence of significant decrease in thought disorder occurrence probability as times passes in hospital; or comparing females to males.

\subsection{British Social Attitudes Panel Survey }

To demonstrate our method for three-level clustered data, we analyzed the \textit{British Social Attitudes Panel Survey} conducted from 1983 through 1986. In this survey, subjects were asked whether they thought there should be no legal or governmental regulation on abortion. This survey was carried out in 54 districts annually for four years among the same individuals. The dataset includes people who have completed all four surveys during the four years, adding up to 1,056 observations from 264 individuals in total. Covariates can be categorized into three levels: the first level is a district-level covariate: the percent of protestants of each district; the second level includes individual-level demographic covariates, including social class (middle, upper and lower), gender (male and female) and religion (Protestant, Catholic, other and none); in the third levels are three dummy variables for years $1984, 1985, 1986$. We can see there are two covariates corresponding to protestant in the model, one on the district-level and the other on the individual-level. By this arrangement we are able to estimate the effect of protestant religion both within district and between districts, as discussed in \cite{ref32}. The inclusion of the two protestant variables are potentially of substantive interest by measuring the religious context or environment impact on individual attitude in contrast to their own religious affiliation affect, as discussed in \cite{ref12}.

In Table 5, point estimates and 95\% confidence intervals of odds ratio corresponding to the above covariates are listed, from three methods. Method 1 is our proposed method assuming the correlation structure of random effects within individuals is auto-regressive of order one and the correlation structure across individuals within a district is exchangeable. Method 2 is also carried out by our proposed method, but assuming both correlation structures within districts and individuals are exchangeable. Method 3 is GEE with an exchangeable working correlation matrix for observations within a district. This ignores the finer level of correlation between observations within individuals, by assuming correlations being equal both within an individual and between two individuals from the same district. We did not compare the results with the MLE as the algorithm failed to converge.

Method 1 and Method 2 give out roughly the same point estimates as GEE but with narrower 95\% confidence intervals. The exceptions are categorical covariates representing religion contrasts between other religions and Protestants within districts having similar proportions of Protestants. This can be explained by the relatively small sample size of this subgroup. In total, we only have 45 individuals of other religions.

Comparing Methods 1 and 2, we can see that the results are roughly the same, indicating the robustness of the proposed method with respect to different assumed correlation structures. Since Method 1 assumes an auto-regressive correlation structure on the third level, where observations are taken annually and therefore their correlations can be better described by an auto-regressive correlation structure
. In the following we report results from Method 1.

The covariates we put into Method 1 decompose religion contrasts into within-cluster contrast and between-cluster contrast. The variable \%Protestant is a district-level covariate and equals to the sampled proportion of district Protestants. The estimated odds ratio is 2.17, 95\% CI: (0.86, 5.52), indicating a non-significantly increasing trend of allowing abortions among individuals from districts of a higher level in Protestants, controlling for all the other variables. The categorical religion contrasts Catholic, other, none, to the reference Protestant group can then be interpreted as comparing the propensity of allowing abortion among individuals of different religions who reside in districts of equal level in Protestants, controlling for year surveyed, social class and gender. Non-significantly lower odds are observed among Catholics in contrast to Protestants with a ratio of 0.67, 95\% CI: (0.25, 1.80), non-significantly lower odds are observed among those of other religions with the ratio 0.52, 95\% CI: (0.24, 1.11) and significantly higher odds are observed among those without any religions with odds ratio being 2.00, 95\% CI: (1.21, 3.30). The propensity of allowing abortions among females is non-significantly lower than that among males from districts of equal level in Protestants, controlling for working class, religion and year of survey, with an odds ratio as 0.72 95\% CI: (0.48, 1.07). The odds ratio of allowing abortions from upper working class comparing to middle class is 0.76, 95\% CI: (0.51, 1.14), and the odds ratio comparing lower working class to middle class is 0.80, 95\% CI: (0.54, 1.19), among people from districts of equal level in Protestants, controlling for gender, religion and year of survey. As for time trend in allowing for abortions propensity, there is a significant drop in Year 1984 compared to the previous year with odds 0.66, 95\% CI: (0.49, 0.88), and there are non-significant increments in the following two years, compared to Year 1983.

\section{Concluding remarks}

In this paper we introduce a marginalizable conditional model for analysing clustered binary data. A working generalized linear mixed effect model and a multivariate Gumbel random intercept distribution are proposed, which yield a marginal logistic regression model that has a population-level interpretation.

Unlike most marginal models which model the first and perhaps the second moment, we have come up with a parametric marginal model, which guarantees there is always a real joint distribution for the marginal logistic regression model and parameters being estimated always exist. In contrast, one criticism of GEE with a cluster-common working correlation matrix for a binary outcome is that there may not be any multivariate distribution with a correlation structure being equivalent to GEE's working correlation structure. 

By generalizing the estimating equation from alternating logistic regression proposed by \cite{ref6}, our proposed inference yields consistent estimates of marginal parameters even under misspecified conditional model or random effect distribution, along with consistent estimates of estimators standard deviation.

The marginalization property is based on a standard exponential frailty assumption, which can be viewed as a special case of the Gamma frailty models considered in \cite{ref18} and \cite{ref22}. However for more general Gamma distributions, the marginal model is no longer conveniently interpretable. Exponential distributed frailties should not be considered as a limitation, since
\begin{enumerate}
\item a marginal logistic model interpretation is often desirable in practice;
\item an exponential distributed frailty is equivalent to a Gumbel random intercept which has physical interpretations. Gumbel distribution can model the distribution of maximum of the normal or exponential type random variables, so Gumbel random intercept is reasonable when we believe there are many latent cluster effects and the maximum dominates the others; i.e. the random effect can be modeled as the maximum of many cluster effects;
\item robust estimation procedure being proposed would yield consistent estimates for marginal parameters even when the multivariate exponential frailty distribution or the conditional mean model is misspecified.
\item marginal inference is un-affected when frailty distribution is covariate dependent.
\end{enumerate}

In this paper we have concentrated on correlated binary outcomes. In principle, our model can be generalized into the cases of correlated ordinal and censored survival data. Investigations are being carried on along these directions.

\section*{Acknowledgment}
The authors thank Professor Jon A. Wellner for his helpful comments on the proof of the theoretical results.

\section*{Appendix}
Here we list conditions of Theorem 1 and prove it. The proof of Theorem 2 is very similar and is omitted.

\begin{eqnarray*}
&&\text{We define }\Psi (\theta)=\left(\ba{c}\Psi_1(\theta)\\ \Psi_2(\theta)\ea\right)=\left(\ba{c}\mathrm{E}\left\{ f_1(X,Y;\theta)\right\}\\ \mathrm{E}\left\{ f_2(X,Y;\theta)\right\}\ea\right)\ ,
\\
&&\text{and }\Psi_m (\theta)=\left(\ba{c}\Psi_{1,m}(\theta)\\ \Psi_{2,m}(\theta)\ea\right)=\left(\ba{c}
\frac{1}{m}\underset{i=1}{\overset{m}{\sum}}
 f_1(X_i,Y_i;\theta)\\ \frac{1}{m}\underset{i=1}{\overset{m}{\sum}} f_2(X_i,Y_i;\theta)\ea\right)\ ,
\end{eqnarray*}
where $f_1$ and $f_2$ correspond to estimating equations in \eqref{EE1} and \eqref{EE2}:
\begin{eqnarray*}
&&\left\{\ba{c}f_1(X_i,Y_i;\theta):=
D(X_i;\beta)^T V^{-1}(X_i;\theta)S(X_i,Y_i;\beta)\ ,\\
f_2(X_i,Y_i;\theta):={\underset{j<k}{\sum}}\frac{\partial l_{jk}}{\partial \rho}(X_i,Y_i;\theta)\ .
\ea\right.
\end{eqnarray*}
Theorem 1 is true under the following conditions:

\noindent C.1 Observations from different clusters are independent and identically distributed.\\
C.2 Number of observations per cluster is uniformly bounded.\\
C.3 Parameter space $\Theta$ is a convex and compact subset of $\mathbb{R}^p$ and the true value of parameter, $\theta_0$, is not a boundary point of $\Theta$.\\
C.4 The probability of covariate $X$ being degenerate is 0, i.e.,  $X^T\beta=0$ a.e. implies $\beta=0$ a.e., and $X$ is bounded with probability one.\\
C.5 There is an unique root of $\beta$ from $\Psi_1(\beta,\rho)=0$ for all $\rho$.\\
C.6 The joint distribution is correctly specified. 
\begin{proof}
First we would like to point out that even though different clusters may contain different numbers of observations, we can still view the joint observations from a cluster as independent and identically distributed (i.i.d).\\
We can regard each cluster in theory contains infinite subjects and their quantities are denoted by $(X(\cdot),Y(\cdot),a(\cdot))$: $\cdot$ varies with different subjects. The data we observe from a cluster is a deterministic projection of $(X(\cdot),Y(\cdot),a(\cdot))$. Assuming the stochastic process $(X(\cdot),Y(\cdot),a(\cdot))$ are i.i.d. and the projection procedure is also i.i.d., we conclude observations from different clusters are i.i.d.. We denote $P_0$ as the joint distribution.

Second, we would like to argue that $\rho_0$ is the unique solution to $\Psi_2(\theta)=0$ at $\beta=\beta_0$. This can be shown by the Kullback-Leibler divergence for composite likelihood.

Composite likelihood of the $i^{th}$ cluster is
\[
{\underset{j<k}{\prod}} L_{jk}(X_i,Y_i;\theta)
\]
The Kullback-Leibler divergence for composite likelihood is
\[
KL_{\text{composite}}(L_0,L_1):=P_0 log \left(\frac{{\underset{j<k}{\prod}}L_0(X_j,X_k,Y_j,Y_k;\beta_0,\rho_0)}{{\underset{j<k}{\prod}}L_1(X_j,X_k,Y_j,Y_k;\beta_0,\rho_1)}\right)
={\underset{j<k}{\sum}}P_0 log\left(\frac{L_{jk}(X_i,Y_i;\beta_0,\rho_0)}{L_{jk}(X_i,Y_i;\beta_0,\rho_1)}\right)>0
\]
the last strict in-equality is due to Jensen's Inequality and the fact that $L_1=L_0$ if and only if $\rho_1=\rho_0$.

Thus $\rho_0$ is the unique value maximizing composite likelihood expectation with plug-in $\beta_0$. Since the model is smooth in parameters, $\Psi_2(\theta)=0$ uniquely at $\rho=\rho_0$ when $\beta$ is fixed at $\beta_0$.

Next, consider an index set $\mathcal{H}:=\{h\in\mathbb{R}^p:||h||\leq 1;\}$ in which $||\cdot||$ is the Euclidean norm. Then the following function class indexed by $\theta\in\Theta$ and $h\in\mathcal{H}$, defined on the sample space of $(X,Y)$, i.e. cluster observations:
\[
\mathcal{F}_0:=\left\{h^T(f_1(X,Y;\theta),f_2(X,Y;\theta)): \theta\in\Theta ,h\in\mathcal{H},(X,Y)\sim P_0
\right\}
\]
is $P_0$-Donsker. 

For an arbitrary pair of functions from $\mathcal{F}_0$:
\begin{eqnarray}
&&|h_1^T(f_1(X,Y;\theta_1),f_2(X,Y;\theta_1))-h_2^T(f_1(X,Y;\theta_2),f_2(X,Y;\theta_2))|\nonumber\\ \leq &&
C_0||\theta_1-\theta_2||\cdot || h_1-h_2 ||
\label{appendix_consistent_jia1}
\end{eqnarray}
This is due to the fact that everything in $h^T(f_1(X,Y;\theta),f_2(X,Y;\theta))$ is continuous in $\theta$ so Mean Value Theorem can be used based on conditions C.2 and C.3; $C_0$ is some finite number by condition C.4.

Since $\theta_1,\theta_2\in\Theta$ and $\Theta$ is a compact subset of Euclidean space, number of brackets needed to cover $\mathcal{F}_0$ satisfies $P_0$-Donsker requirement, according to \cite{book5}, page 129.

Now we can claim
\[
{\underset{\theta\in\Theta,h\in\mathcal{H}}{\text{sup}}}|h^T\Psi_m(\theta)-h^T\Psi(\theta)|\rightarrow 0
\]
implying that
\begin{eqnarray*}
&&{\underset{h\in\mathcal{H}}{\text{sup}}}|h^T\left[\Psi_m(\hat{\theta}_m)-\Psi(\hat{\theta}_m)\right]|\rightarrow 0\ ,\\ \text{i.e. }&&
{\underset{h\in\mathcal{H}}{\text{sup}}}|h^T\Psi(\hat{\theta}_m)|\rightarrow 0\ ;\quad \text{thus, } |\Psi(\hat{\theta}_m)|\rightarrow 0\ .
\end{eqnarray*}
Since $(f_1(X,Y;\theta),f_2(X,Y;\theta))$ are continuous in $\theta$, we have shown $\hat{\theta}_m{\overset{p.}{\rightarrow}} \theta_0$.
\end{proof}

The proof of the weak convergence of $\sqrt{m}\left\{(\hat{\beta}_m-\beta_0)^T,(\hat{\rho}_m-\rho_0)^T\right\}^T$ makes use of Theorem 3.3.1 of \cite{book5}, which is stated as the following.\\
Suppose there are two random mappings $\Psi_m$ and $\Psi$ such that $\Psi(\beta_0,\rho_0)=0$ 
for some interior point $(\beta_0,\rho_0)\in\Theta$, $\Psi_m(\beta_m,\rho_m){\overset{p.}{\rightarrow}}0$ for some random sequence $(\beta_m,\rho_m)\subset \Theta$,
and assume the followings are true:\\
P.1 $(\beta_m,\rho_m)$ is consistent for $(\beta_0,\rho_0)$;\\
P.2 $\sqrt{m}\left(\Psi_m-\Psi\right)(\beta_0,\rho_0)$ converges in distribution to a tight random element $Z$;\\
P.3 
\begin{eqnarray*}
&&\sqrt{m}\left(\Psi_m-\Psi\right)(\beta_m,\rho_m)-
\sqrt{m}\left(\Psi_m-\Psi\right)(\beta_0,\rho_0)\\=&&o_p\left(
1+\sqrt{m}||\beta_m-\beta_0||+\sqrt{m}||\rho_m-\rho_0||
\right)\ ;
\end{eqnarray*}
P.4 $\Psi(\beta,\rho)$ is Fr\'{e}chet differentiable at $(\beta_0,\rho_0)$;\\
P.5 The derivative of $\Psi(\beta,\rho)$ at $(\beta_0,\rho_0)$, denoted by $\dot{\Psi}(\beta_0,\rho_0)$, is continuously invertible.

Then 
\[
\sqrt{m}\left\{(\hat{\beta}_m-\beta_0)^T,(\hat{\rho}_m-\rho_0)^T\right\}^T{\overset{d.}{\rightarrow}}-\dot{\Psi}(\beta_0,\rho_0)^{-1}(Z) \ .
\]
\begin{proof}
Condition P.1 has been verified.
 
Since we have shown $\mathcal{F}_0$ is $P_0$-Donsker, condition P.2 is verified.

By $P_0$-Donsker preservation theorem 2.10.3 in \cite{book5}, this function class
\[
\left\{
h^T\left[\left\{(f_1(\beta,\rho),f_2(\beta,\rho))-(f_1(\beta_0,\rho_0),f_2(\beta_0,\rho_0))\right\}\right]:(\beta,\rho)\in \Theta, h\in\mathcal{H}\right\}
\]
is $P_0$-Donsker as well.
\begin{eqnarray*}
&&{\underset{h\in\mathcal{H}}{\text{sup}}}P_0\left(h^T\left[\left\{(f_1(\beta,\rho),f_2(\beta,\rho))-(f_1(\beta_0,\rho_0),f_2(\beta_0,\rho_0))\right\}\right]\right)^2\nonumber\\ \leq && P_0\left(C_0 ||\theta_0-\theta||\right)^2
\rightarrow   0\quad\text{ as }||(\beta,\rho)-(\beta_0,\rho_0)||\rightarrow 0\nonumber
\end{eqnarray*}
Therefore, according to Lemma 3.3.5 of \cite{book5}, P.3 holds.

As for P.4, since $\Psi$ is a smooth function in parameters, it is trivial to verify that $-E(B)$ is its Fr\'{e}chet derivative at $(\beta_0,\rho_0)$. Due to model identifiability and condition C.3, $E(B)$ is a negative definite matrix and thus continuously invertible. Therefore, P.5 is also satisfied.
\end{proof}

\bibliographystyle{apalike}
\bibliography{Refs1}		

\begin{table}[h]
\begin{center}
\caption{\sl{Simulation results for estimating $(\beta_0,\beta_1,\rho_0)$ in two-level clustering, where $\rho_0$ is the correlation parameter of random effects. Bias represents the empirical bias, SSE represents the Monte Carlo standard error (s.e.), MSE is the mean squared error. SEE represents the averaged model-based s.e. estimates.\label{T1}}}
\medskip
\begin{small}
\begin{tabular}{|c|c|cc|cc|cc|cc|cc|c|c|} \hline
\multicolumn{14}{|c|}{(a) Two-level clustering, exchangeable correlation matrix.}\\ \hline
$\rho_0$&Method&\multicolumn{2}{|c|}{Bias$\times 10^{3}$}&\multicolumn{2}{c|}{SEE$\times 10^{3}$}&\multicolumn{2}{c|}{SSE$\times 10^{3}$}&\multicolumn{2}{c|}{MSE$\times 10^{3}$}&\multicolumn{2}{c|}{95\% C.I. coverage rate}&Bias$\times 10^{3}$&Computing\\
&&$\hat{\beta}_0$&$\hat{\beta}_1$&$\hat{\beta}_0$&
$\hat{\beta}_1$&$\hat{\beta}_0$&$\hat{\beta}_1$&
$\hat{\beta}_0$&$\hat{\beta}_1$&$\hat{\beta}_0$&
$\hat{\beta}_1$&$\hat{\rho}_0$&Times (sec)\\ \hline
\multirow{2}{*}{0.1}&Proposed&-1&3&93&71&93&69&9&5&96.0\%&96.2\%&6&5\\
&MLE&-1&3&93&71&93&69&9&5&96.0\%&96.2\%&2&44\\
\multirow{2}{*}{0.3}&Proposed&-5&7&98&71&103&73&11&5&93.6\%&95.4\%&-16&6\\
&MLE&-5&7&98&71&103&73&11&5&93.7\%&95.4\%&-16&51\\
\multirow{2}{*}{0.5}&Proposed&-6&2&104&71&106&69&11&5&94.5\%&96.0\%&-13&6\\
&MLE&-6&2&104&71&106&69&11&5&94.5\%&96.0\%&-12&53\\
\multirow{2}{*}{0.7}&Proposed&-8&2&112&72&114&71&13&5&94.9\%&95.0\%&-11&6\\
&MLE&-9&2&112&72&113&71&13&5&94.4\%&95.0\%&-9&51\\
\multirow{2}{*}{0.9}&Proposed&-9&7&123&75&121&72&15&5&94.4\%&96.6\%&-9&6\\
&MLE&68&-47&117&69&111&63&17&6&91.4\%&89.4\%&71&35\\ \hline
\end{tabular}
\vspace{80pt}
\begin{tabular}{|c|c|cc|cc|cc|cc|cc|c|c|} \hline
\multicolumn{14}{|c|}{(b) Two-level clustering, AR(1) correlation matrix.}\\ \hline
$\rho_0$&Method&\multicolumn{2}{|c|}{Bias$\times 10^{3}$}&\multicolumn{2}{c|}{SEE$\times 10^{3}$}&\multicolumn{2}{c|}{SSE$\times 10^{3}$}&\multicolumn{2}{c|}{MSE$\times 10^{3}$}&\multicolumn{2}{c|}{95\% C.I. coverage rate}&Bias$\times 10^{3}$&Computing\\
&&$\hat{\beta}_0$&$\hat{\beta}_1$&$\hat{\beta}_0$&
$\hat{\beta}_1$&$\hat{\beta}_0$&$\hat{\beta}_1$&
$\hat{\beta}_0$&$\hat{\beta}_1$&$\hat{\beta}_0$&
$\hat{\beta}_1$&$\hat{\rho}_0$&Times (sec)\\ \hline
\multirow{2}{*}{0.1}&Proposed&-4&6&92&71&88&72&8&5&95.6\%&95.1\%&30&5\\
&MLE&-4&6&92&71&88&72&8&5.2&95.6\%&95.0\%&36&75\\
\multirow{2}{*}{0.3}&Proposed&-9&11&94&71&95&74&9&6&94.0\%&94.0\%&-25&7\\
&MLE&-9&11&94&71&95&74&9&6&94.1\%&94.2\%&-16&63\\
\multirow{2}{*}{0.5}&Proposed&-7&6&98&71&106&72&11&5&93.8\%&94.6\%&-20&7\\
&MLE&-7&66&98&71&106&72&11&5&93.6\%&94.7\%&-17&62\\
\multirow{2}{*}{0.7}&Proposed&-8&9&104&72&106&73&11&5&94.6\%&94.8\%&-16&9\\
&MLE&-9&9&104&72&106&73&11&5&94.4\%&95.0\%&-11&63\\
\multirow{2}{*}{0.9}&Proposed&-6&8&114&73&118&71&14&5&94.3\%&95\%&-7&31\\
&MLE&134&-89&111&66&130&80&35&14.3&72.3\%&63.8\%&-14&35\\ \hline
\end{tabular}
\end{small}
\end{center}
\end{table}

\begin{table}[h]
\begin{center}
\caption{\sl{Simulation results for estimating $(\beta_0,\beta_1,\rho_2,\rho_3)$ in a three-level clustering. We assume exchangeable correlation structure in both levels of clustering, and $(\rho_2,\rho_3)$ represents the true correlations in the second and the third clustering levels. Bias, SSE, SEE, MSE represent the same quantities as in Table 1. 95\% confidence interval coverage rates are presented, derived from model based s.e..\label{T2}}}
\begin{small}
\begin{tabular}{|cc|c|cc|cc|cc|cc|cc|cc|c|} \hline
\multicolumn{16}{|c|}{Three-level clustering, exchangeable correlation matrix.}\\ \hline
$\rho_2$&$\rho_3$&Method&\multicolumn{2}{|c|}{Bias}&\multicolumn{2}{c|}{SEE}&\multicolumn{2}{c|}{SSE}&\multicolumn{2}{c|}{MSE}&\multicolumn{2}{c|}{95\% C.I.}&\multicolumn{2}{c|}{Bias}&Computing\\
&&&\multicolumn{2}{c|}{$\times 10^{3}$}&\multicolumn{2}{c|}{$\times 10^{3}$}&\multicolumn{2}{c|}{$\times 10^{3}$}&\multicolumn{2}{c|}{$\times 10^{3}$}&\multicolumn{2}{c|}{ coverage rate}&\multicolumn{2}{c|}{$\times 10^{3}$}&Times (sec)\\
&&&$\hat{\beta}_0$&$\hat{\beta}_1$&$\hat{\beta}_0$&
$\hat{\beta}_1$&$\hat{\beta}_0$&$\hat{\beta}_1$&
$\hat{\beta}_0$&$\hat{\beta}_1$&$\hat{\beta}_0$&
$\hat{\beta}_1$&$\hat{\rho}_2$&$\hat{\rho}_3$&\\ \hline
\multirow{2}{*}{0.1}&\multirow{2}{*}{0.1}&Proposed&-4&9&84&121&85&124&7&15&94.7\%&95.1\%&13&53&36\\
&&MLE&-4&10&80&121&85&124&7&16&93.6\%&95.3\%&-68&-61&57\\
\multirow{2}{*}{0.1}&\multirow{2}{*}{0.3}&Proposed&$<1$&2&86&120&88&123&8&15&94.3\%&95.2\%&18&10&23\\
&&MLE&$<1$&2&82&121&89&124&8&15&93.2\%&95.2\%&-27&-184&72\\
\multirow{2}{*}{0.1}&\multirow{2}{*}{0.5}&Proposed&5&6&89&120&89&123&8&15&95.6\%&94.8\%&15&17&21\\
&&MLE&2&-2&85&120&93&127&9&16&93.4\%&94.3\%&-12&-247&78\\
\multirow{2}{*}{0.1}&\multirow{2}{*}{0.7}&Proposed&-1&6&93&118&90&127&8&16&95.6\%&93.8\%&15&21&20\\
&&MLE&41&-28&89&114&96&123&11&16&90.3\%&91.0\%&-33&85&59\\
\multirow{2}{*}{0.3}&\multirow{2}{*}{0.1}&Proposed&-4&4&89&120&92&117&8&14&95.1\%&95.6\%&-12&32&19\\
&&MLE&-4&5&85&121&92&118&9&14&93.8\%&95.3\%&-155&11&72\\
\multirow{2}{*}{0.3}&\multirow{2}{*}{0.3}&Proposed&-5&6&91&120&91&121&8&15&95.6\%&94.7\%&-17&6&9\\
&&MLE&-4&5&90&120&92&121&9&15&94.4\%&94.6\%&-48&-17&96\\
\multirow{2}{*}{0.3}&\multirow{2}{*}{0.5}&Proposed&-6&7&95&119&95&118&9&14&94.6\%&95.4\%&-13&5&7\\
&&MLE&1&1&94&118&97&118&9&14&93.4\%&95.2\%&-43&36&98\\
\multirow{2}{*}{0.5}&\multirow{2}{*}{0.1}&Proposed& -6&3&97&119&98&120&10&14&95.2\%&95.2\%&-8&17&15\\
&&MLE&13&-10&91&118&118&134&14&18&88.2\%&91.0\%&-150&7&75\\
\multirow{2}{*}{0.5}&\multirow{2}{*}{0.3}&Proposed&-6&2&99&118&104&119&11&14&93.6\%&94.7\%&-16&4&6\\
&&MLE&3&-6&98&117&110&122&12&15&92.1\%&93.1\%&-56&50&97\\
\multirow{2}{*}{0.7}&\multirow{2}{*}{0.1}&Proposed&-8&4&105&118&109&123&12&15&93.0\%&93.5\%&-11&5&13\\
&&MLE&97&-89&102&107&126&123&25&23&78.6\%&79.5\%&90&6&63 \\ \hline
\end{tabular}
\end{small}
\end{center}
\end{table}

\begin{table}[h]
\begin{center}
\caption{\sl{Simulation results for estimating $(\beta_0,\beta_1)$ in a two-level clustering setting with a misspecified joint model but a correct marginal model. Bias, SSE, SEE, MSE represent the same quantities as in Table 1. Two SEE's are presented, one is model-based while the other is robust. 95\% confidence interval coverage rates are presented, derived by model based s.e. and robust s.e., respectively.\label{T3}}}
\begin{small}
\begin{tabular}{|c|cc|cc|cc|cc|cc|cc|cc|} \hline
\multicolumn{15}{|c|}{Two-level clustering, mis-specified conditional model.} \\ \hline
Method&\multicolumn{2}{|c|}{Bias}&\multicolumn{2}{c|}{SEE}&\multicolumn{2}{c|}{Robust SEE}&\multicolumn{2}{c|}{SSE}&\multicolumn{2}{c|}{MSE}&\multicolumn{2}{c|}{95\% C.I.}&\multicolumn{2}{c|}{Robust 95\% C.I.}\\
&\multicolumn{2}{|c|}{$\times 10^{3}$}&\multicolumn{2}{c|}{$\times 10^{3}$}&\multicolumn{2}{c|}{$\times 10^{3}$}&\multicolumn{2}{c|}{$\times 10^{3}$}&\multicolumn{2}{c|}{$\times 10^{3}$}&\multicolumn{2}{c|}{coverage rate}&\multicolumn{2}{c|}{coverage rate}\\
&$\hat{\beta}_0$&$\hat{\beta}_1$&$\hat{\beta}_0$&$\hat{\beta}_1$&$\hat{\beta}_0$&$\hat{\beta}_1$&$\hat{\beta}_0$&$\hat{\beta}_1$&$\hat{\beta}_0$&$\hat{\beta}_1$&$\hat{\beta}_0$&$\hat{\beta}_1$&$\hat{\beta}_0$&$\hat{\beta}_1$\\ \hline
Proposed&9&16&133&150&153&151&155&148&24&22&91.1\%&95.5\%&94.8\%&95.9\%\\
MLE&-308&201&144&164&-100&-100&159&153&121&64&42.0\%&81.4\%&-&-\\ \hline
\end{tabular}
\end{small}
\end{center}
\end{table}

\begin{table}[h]
\begin{center}
\caption{\sl{Analysis of \textit{Madras longitudinal schizophrenia study.} \label{T4}}}
\medskip
\begin{tabular}{l|cccccc}
\hline
&&\multicolumn{2}{c}{Exchangeable}&&\multicolumn{2}{c}{AR (1)}\\
\cline{3-4} \cline{6-7}
Coefficients&&$\exp(\beta)$&95\% C.I.&&$\exp(\beta)$&95\% C.I.\\ \hline
\\&&\multicolumn{5}{c}{Likelihood}\\ \\
Intercept&&2.29&(1.44, 3.66)&&2.27&(1.43, 3.61)\\
Time&&0.70&(0.66, 0.75)&&0.70&(0.66, 0.75)\\
Age $\leq 20$&&1.50&(0.83, 2.72)&&1.31&(0.73, 2.34)\\
Female&&0.43&(0.24, 0.79)&&0.45&(0.26, 0.79)\\ 
$\rho$&&\multicolumn{2}{c}{0.94} &&\multicolumn{2}{c}{0.95}\\ 
\\&&\multicolumn{5}{c}{Proposed Method}\\ \\
Intercept&&2.41&(1.54, 3.78)&&2.49&(1.57, 3.93)\\
Time&&0.71&(0.67, 0.75)&&0.71&(0.67, 0.76) \\
Age $\leq 20$&&1.60&(0.88, 2.90)&&1.47&(0.81, 2.66)\\
Female&&0.53&(0.30, 0.95)&&0.54&(0.30, 0.96)\\ 
$\rho$&&\multicolumn{2}{c}{0.92} &&\multicolumn{2}{c}{0.96}\\ \hline
\end{tabular}
\end{center}
\end{table}
\begin{table}[h]
\begin{center}
\caption{\sl{Analysis of \textit{British Social Attitudes Panel Survey: years 1983-1986.} \label{T5}}}
\medskip
\begin{tabular}{l|cccccccc}
\hline
&\multicolumn{2}{c}{Method 1}&&\multicolumn{2}{c}{Method 2}&&\multicolumn{2}{c}{Method 3}\\  \cline{2-3} \cline{5-6} \cline{8-9}
Coefficients&$\exp(\beta)$&95\% C.I.&&$\exp(\beta)$&95\% C.I.&&$\exp(\beta)$&95\% C.I.\\ \hline
Intercept&0.61&(0.23, 1.63)&&0.62&(0.24, 1.61)&&0.74&(0.22, 2.43)\\[1.5ex]
Year 1984&0.66&(0.49, 0.88)&&0.65&(0.48, 0.88)&&0.65&(0.47, 0.91)\\[1.5ex]
Year 1985&1.06&(0.80, 1.41)&&1.05&(0.78, 1.40)&&1.04&(0.74, 1.46)\\ [1.5ex]
Year 1986&1.21&(0.91, 1.61)&&1.20&(0.90, 1.61)&&1.20&(0.88, 1.63)\\[1.5ex]
Class: upper working&0.76&(0.51, 1.14)&&0.75&(0.50, 1.13)&&0.72&(0.41, 1.24)\\[1.5ex]
Class: lower working&0.80&(0.54, 1.19)&&0.78&(0.52, 1.16)&&0.66&(0.43, 1.02)\\[1.5ex]
Gender&0.72&(0.48, 1.07)&&0.72&(0.49, 1.07)&&0.71&(0.45, 1.11)\\[1.5ex]
Religion:  catholic&0.67&(0.25, 1.80)&&0.67&(0.26, 1.76)&&0.76&(0.30, 1.91)\\[1.5ex]
Religion: other&0.52&(0.24, 1.11)&&0.52&(0.25, 1.08)&&0.45&(0.23, 0.87)\\[1.5ex]
Religion: none&2.00&(1.21, 3.30)&&2.02&(1.23, 3.29)&&2.12&(1.13, 3.97)\\[1.5ex]
\% protestant&2.17&(0.86, 5.52)&&2.19&(0.88, 5.48)&&1.94&(0.70, 5.41)\\ \hline
\end{tabular}
\end{center}
\end{table}

\end{document}